\documentclass[runningheads]{llncs}
\usepackage[T1]{fontenc}

\usepackage{multirow}
\usepackage{cite}
\usepackage{amsmath,amssymb,amsfonts}
\usepackage{algorithm,algorithmic}
\usepackage{graphicx}
\usepackage{subfigure}
\usepackage{textcomp}
\usepackage{xcolor}
\usepackage{comment}
\usepackage{multirow}
\usepackage{booktabs}
\usepackage{array}
\usepackage{algorithm, algorithmic}
\usepackage{pgf}
\usepackage{hyperref}
\usepackage{makecell}
\usepackage{color}
\usepackage{fancyhdr}

\begin{document}
\title{A study on the adequacy of common IQA measures for medical images}
\titlerunning{Adequacy of common IQA measures for medical images}

\author{Anna Breger \inst{1,2}$^*$
    \and Clemens Karner \inst{2}
    \and Ian Selby \inst{3,4}
    \and Janek Gröhl \inst{5,6}
    \and Sören Dittmer \inst{1} 
    \and Edward Lilley \inst{2}
    \and Judith Babar \inst{3,4}
    \and Jake Beckford \inst{4}
    \and Thomas R Else \inst{5,6}
    \and Timothy J Sadler \inst{4}
    \and Shahab Shahipasand \inst{4}
    \and Arthikkaa Thavakumar \inst{4} 
    \and Michael Roberts \inst{1}
    \and Carola-Bibiane Schönlieb \inst{1}
%Anna Breger \inst{1}\orcidID{0000-1111-2222-3333} \and
%Clemens Karner\inst{2,3}\orcidID{1111-2222-3333-4444} \and
%Third Author\inst{3}\orcidID{2222--3333-4444-5555}
}
\authorrunning{A. Breger et al.}
% First names are abbreviated in the running head.
% If there are more than two authors, 'et al.' is used.
%
\institute{University of Cambridge, DAMTP, Cambridge, United Kingdom \and Medical University of Vienna, CMPBE, Vienna, Austria \and University of Cambridge, Department of Radiology, Cambridge, UK \and Cambridge University Hospitals, Department of Radiology, Cambridge, UK \and University of Cambridge, Department of Physics, Cambridge, UK \\ \and Cancer Research UK Cambridge Institute, Cambridge, UK \\
\email{$^*$Corresponding author. E-mail: ab2864@cam.ac.uk}}

\maketitle              % typeset the header of the contribution

\begin{abstract}
Image quality assessment (IQA) is standard practice in the development stage of novel machine learning algorithms that operate on images. The most commonly used IQA measures have been developed and tested for natural images, but not in the medical setting. Reported inconsistencies arising in medical images are not surprising, as they have different properties than natural images. In this study, we test the applicability of common IQA measures for medical image data by comparing their assessment to manually rated chest X-ray (5 experts) and photoacoustic image data (2 experts). Moreover, we include supplementary studies on grayscale natural images and accelerated brain MRI data. The results of all experiments show a similar outcome in line with previous findings for medical images: PSNR and SSIM in the default setting are in the lower range of the result list and HaarPSI outperforms the other tested measures in the overall performance. Also among the top performers in our experiments are the full reference measures FSIM, LPIPS and MS-SSIM. Generally, the results on natural images yield considerably higher correlations, suggesting that additional employment of tailored IQA measures for medical imaging algorithms is needed. 

\end{abstract}

\section{Introduction}
Advances in medical imaging technologies have been groundbreaking in the last decades, including the rapid development of deep learning techniques. To ensure the quality of novel image processing methodologies, quantitative image quality assessment (IQA) plays an important role for quality assurance in addition to visual inspection. It may even serve as the main assessment criterion when no expert opinion is available. Quantitative IQA measures can be distinguished based on the needed information for the evaluation step \cite{ssim, perfcomp}. In full reference (FR) IQA a reference image is used to evaluate the quality of a corresponding (often degraded) image in a comparative way, relying on a meaningful notion of distance between the two images. No reference (NR) IQA aims to judge the quality without a reference based on pre-defined properties. Reduced reference (RR) IQA uses derived image information of reference data. \\
Most common IQA measures have been developed for natural images and tested for specific tasks on a small amount of publicly available, manually rated data sets. It is unknown how well these measures expand to medical images since they have very distinct properties, and, moreover, a different target space (color versus grayscale). Little research has been performed on the applicability of common IQA measures to medical imaging data, see e.g.~\cite{iqamri} for a recent overview. Many prior applicability studies have limitations in the study design, including non-expert ratings (see e.g.~\cite{mrinonexpert}), non-realistic distortions (such as Gaussian additive noise, see e.g.~\cite{gaussct}) and a very limited choice of IQA measures (see e.g.~\cite{ssimcompmri}). The research field suffers from the lack of publicly available data sets with expert annotations, as well as non-available codes of published IQA measures. Combined with the scarce time resources of medical experts, there are many obstacles for the design of reproducible IQA comparison studies. \\ 
Recently, in \cite{iqamri}, a first extensive study on IQA measures comparing MRI outputs of image restoration methods with FR expert ratings has been published. The results suggest that the most widely used FR-IQA measures, namely PSNR \cite{639240} and SSIM \cite{ssim}, are not a good choice for the tested MRI tasks, which is in line with previous findings (see e.g. \cite{iqafail} for an overview of challenges when applying these two measures to medical images). \\ 
Building upon that work, here, we introduce a first study that includes multiple medical image modalities for the comparison of common IQA measures with expert ratings. In particular, we design $4$ experimental setups with $5$ data sets, including expert ratings for photoacoustic and chest X-ray data. The following common IQA measures have been included for comparison: peak signal-to-noise ratio (PSNR), structural similarity index measure (SSIM), multi-scale SSIM (MS-SSIM) \cite{msssim}, information content weighted SSIM (IW-SSIM) \cite{iwssim}, deep image structure and texture similarity (DISTS) \cite{9298952}, DCT subband similarity (DSS) \cite{dss}, feature similarity index (FSIM) \cite{5705575}, gradient magnitude similarity deviation (GMSD) \cite{gmsd}, Haar wavelet-based perceptual similarity (HaarPSI) \cite{Reisenhofer18}, learned perceptual image patch similarity (LPIPS) \cite{lpips}, mean deviation similarity (MDSI) \cite{mdsi}, visual information fidelity (VIF) \cite{vif}, and visual saliency-induced index (VSI) \cite{vsi}, %PieAPP \cite{pieapp}, 
in the FR setting; blind/referenceless image spatial quality evaluator BRISQUE \cite{brisque}, 
from patches to pictures (PaQ-2-PIQ) \cite{paq2piq}, and natural image quality evaluator (NIQE) \cite{niqe} in the NR setting. \\
The experiments are described in detail in Section 2 and results are discussed in Section 3. The data is made available in line with the requirements of each data set, see Section 4. 

\section{Methods}\label{methods}
\begin{figure}
\centering
\includegraphics[width=0.95\textwidth]{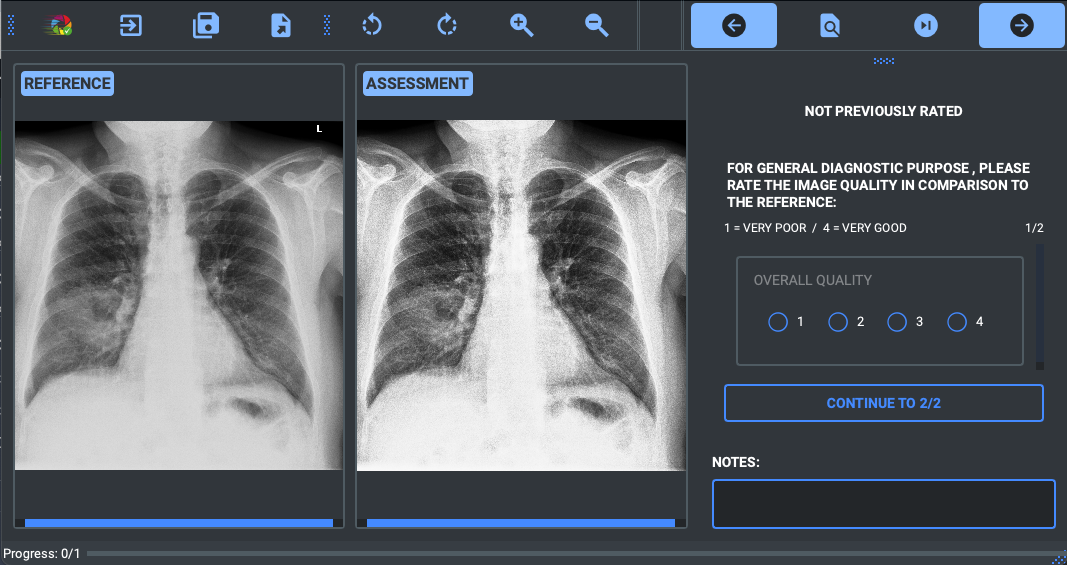}
\caption{The speedyIQA annotation app allows setting a task and rating categories for manual image quality ratings.}
    \label{speedyiqa}
\end{figure}

The IQA measures were computed with the implementations provided by their authors, either in MATLAB or Python, or both. An exception is the VIF measure for which we used the PyTorch Lightning implementation because the code by the authors is not publicly available. Reporting the implementation used is of utmost importance because IQA results may differ substantially in regards to the implementation, see e.g.~\cite{Venkataramanan2021AHG}. \\
When novel manual image ratings were obtained (Experiment $1$, $3$, and $4$), the novel publicly available speedyIQA annotation app \cite{speedyiqa} was used with the uncompressed image data. The software asks the user to set a task and the rating categories, see Figure \ref{speedyiqa}. Obtained ratings have been saved as a CSV file. \\
For the evaluation of the IQ measure performance we employed the Spearman Rank Correlation Coefficient (SRCC) and the Kendall Rank Correlation Coefficient (KRCC), which assess the ordinal association (rank) between the IQ measures and the manual ratings. To account for the different scoring between multiple graders, we apply the z-score to the raw rating data of each grader and afterwards compute the mean score, cf.~\cite{1709988}. For the evaluation, the absolute SRCC and KRCC between the mean rating score and the IQ values are stated. 

\vspace{-1cm}
\subsection{Experiment 1: Grayscale LIVE Data}
The two data sets in the first experiment correspond to often used natural imaging quality assessment databases, namely the LIVE Image Quality Assessment Database Release 2 and the LIVE Multiply Distorted Image Quality Database \cite{live, test}, which we transformed to grayscale images with the in-built MATLAB function \textit{mat2gray} to match the usual predominant target space of medical images. The data sets contain respectively $982$ and $405$ images, including degraded images with Gaussian noise, jpeg compression, and blurring. Five volunteers were asked to rate all degraded images of both data sets from $1$ (very poor), $2$ (poor), $3$ (good) to $4$ (very good) regarding the ability to identify the detailed image content in comparison to the given reference image. In order to compare the results in the same target space, we did not use the original available color image ratings, see e.g.~\cite{Reisenhofer18} for an example of inconsistencies related to grayscale versus color image ratings. 

\vspace{-0.1cm}
\subsection{Experiment 2: MRI Acceleration}
\begin{figure}
%brain_AXT1PRE_210_6001831_3.png
\centering 
\includegraphics[angle=180,origin=c,width=0.16\textwidth]{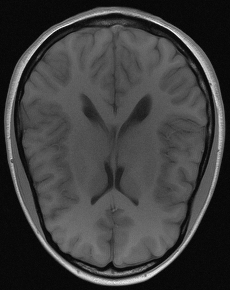}
\includegraphics[angle=180,origin=c,width=0.16\textwidth]{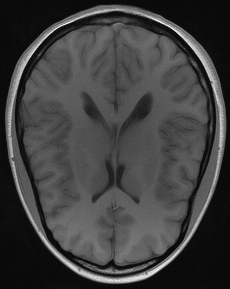} 
\includegraphics[angle=180,origin=c,width=0.16\textwidth]{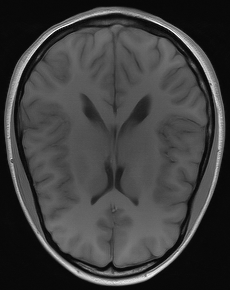} \includegraphics[angle=180,origin=c,width=0.16\textwidth]{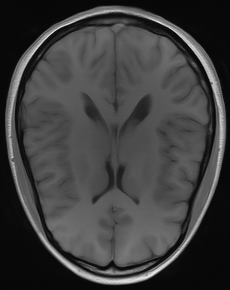} 
\includegraphics[angle=180,origin=c,width=0.16\textwidth]{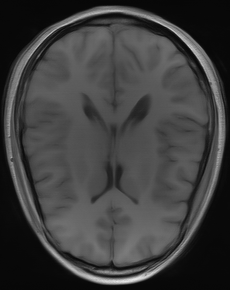} 
\includegraphics[angle=180,origin=c,width=0.16\textwidth]{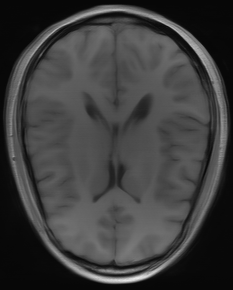} 
\caption{\small Reconstructed MRI brain data from the fastMRI data set obtained with \textit{E2E-VarNet} on the sub-sampled data with acceleration factors $1, 4, 8, 12$ and $16$. The left image corresponds to the reference image obtained via the rSOS of the fully sampled data. The visual quality decreases with the increased acceleration factor.}
\label{mri}
\end{figure}

The MRI data set was retrieved from the publicly available fastMRI brain dataset \cite{fastmri}, which contains in total $6405$ T1, T2, and FLAIR 3D k-space volumes. The fastMRI challenge series provided MRI datasets to foster the development of accelerated reconstruction algorithms. In \cite{iqamri} data from the fastMRI data set has been used for a comprehensive analysis of common IQA measures compared to expert annotations. Here, we use a subset of $4742$ reference image slices (created by the root sum of squares, rSOS, of the fully sampled data) and around $151k$ corresponding accelerated image reconstructions obtained from two machine learning algorithms that took part in the fastMRI multi-coil brain dataset challenge in 2020, namely the end-to-end variational network \textit{E2E-VarNet} \cite{e2evarnet} and \textit{XPDNet} \cite{xpdnet}. \textit{XPDNet} was among the top three submissions of the challenge and both algorithms perform very well on the corresponding public leaderboard. %\cite{leaderboardfastmri2}. 
Reconstructions were obtained by the application of the machine learning models to the fully sampled and accelerated data, where we designed masks with acceleration factor $1$ to $16$ to yield decreasing visual image quality in the reconstructions, see Figure \ref{mri}. \\ 
The created data set serves as a sanity check for the identification of decreasing image quality. We evaluate the performance of the IQA measures in two ways: First, the SRCC and KRCC are computed for each image and the corresponding quality decreases, where the acceleration factor serves as the image quality category, and secondly, we plot the mean IQA value for each acceleration class and measure. For this illustration, all measures are linearly scaled such that a lower value is worse, starting with the IQA value of acceleration factor $1$ as $100\%$. The purpose of this experiment is to test the IQA measure's ability to flawlessly detect distinct quality decrease in medical images. 

\vspace{-0.1cm}
\subsection{Experiment 3: Photoacoustic Reconstruction}
\begin{figure}[h!]
\centering 
\includegraphics[width=0.21\textwidth]{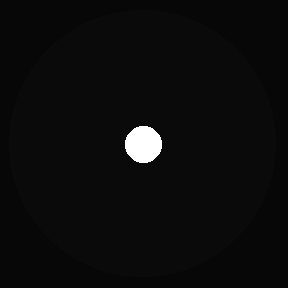} \ \
\includegraphics[width=0.21\textwidth]{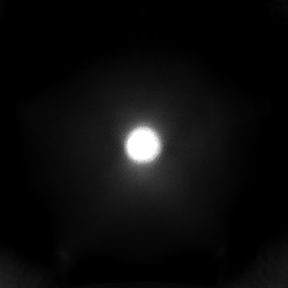} \ \
\includegraphics[width=0.21\textwidth]{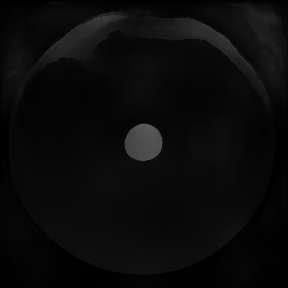} \ \
\includegraphics[width=0.21\textwidth]{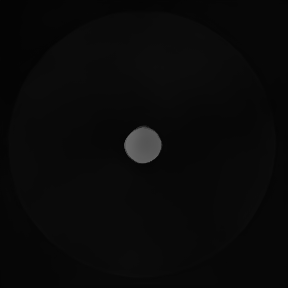} \\
\subfigure[Reference]{
\includegraphics[width=0.21\textwidth]{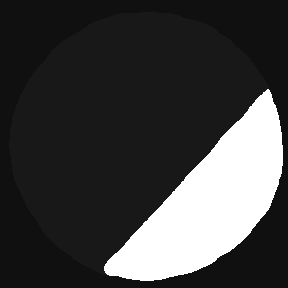} }
\subfigure[Algorithm 1]{
\includegraphics[width=0.21\textwidth]{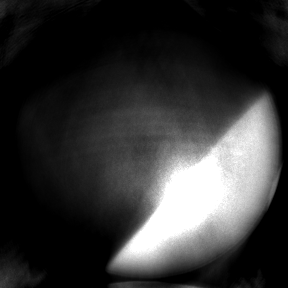} }
\subfigure[Algorithm 2]{
\includegraphics[width=0.21\textwidth]{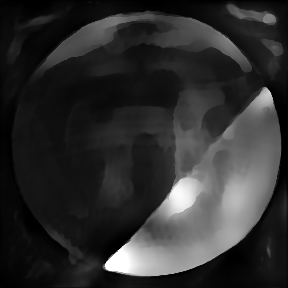} }
\subfigure[Algorithm 3]{
\includegraphics[width=0.21\textwidth]{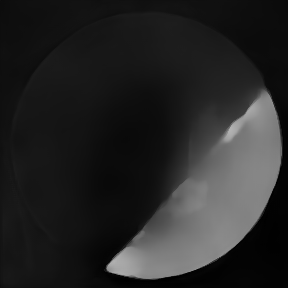} }

\caption{\small Two examples of the photoacoustic images used, references (a) and reconstructions from three algorithms (b)-(d). Algorithm 1 corrects a reconstructed PA image by using the light fluence obtained from simulations. Algorithm 2 and 3 are deep-learning models trained to estimate the absorption coefficient.}
    \label{photoiqa}
\end{figure}

Photoacoustic (PA) imaging is an emerging medical imaging modality with important clinical applications such as inflammatory bowel disease,  cardiovascular diseases, and breast cancer~\cite{ASSI2023100539}. %~\cite{Janek1,Janek2}. 
The inverse problems of PAI pertain to accurately visualizing molecular distributions and determining functional tissue information from PA time series measurements~\cite{Janek5}.
We use a previously published open access data set, cf.~\cite{Janek7}, that consists of reconstructed images containing estimated distributions of the optical absorption coefficient from cross-sectional photoacoustic images of piecewise constant test objects (phantoms). The PA data were acquired with a preclinical commercial photoacoustic imaging system. %(MSOT InVision 256-TF, iThera Medical GmbH, Munich, Germany). 
The $378$ reference images are obtained using a double-integrating sphere \cite{Janek8} setup as a complementary measurement system, which yields point estimates for homogeneous material samples. Because of the piecewise-constant nature of the used phantoms, one can fabricate an additional batch of the material used for the test object, measure it, and relate the calculated properties to the test object. This process is unfeasible for complicated objects or in vivo images, but can serve in this setting to obtain reference images. \\
Here, $1134$ reconstructed images, corresponding to the outputs of $3$ different reconstruction methods (see Figure~\ref{photoiqa}), have been annotated by 2 experts. They were asked to rate the images from $1$ (very poor), $2$ (poor), $3$ (good) to $4$ (very good) regarding overall quality in comparison to the reference image without changing the contrast or luminance. For visualization and assessment, the outputs of the algorithms were clipped with the reference image's maximum. 

\vspace{-0.35cm}
\subsection{Experiment 4: Chest X-Ray Post-Processing}
\begin{figure}
  \centering
\includegraphics[width=0.24\textwidth]{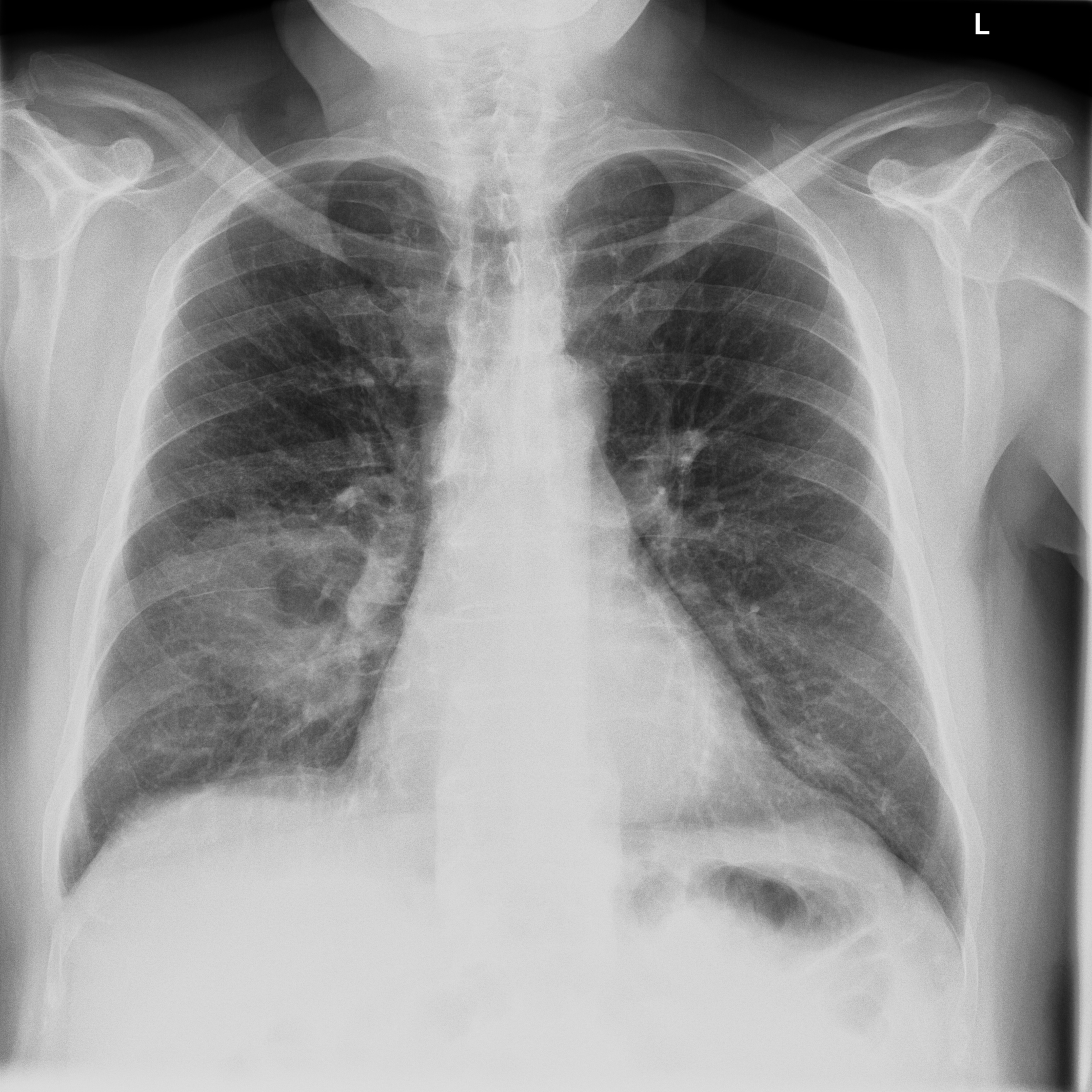} 
  \includegraphics[width=0.24\textwidth]{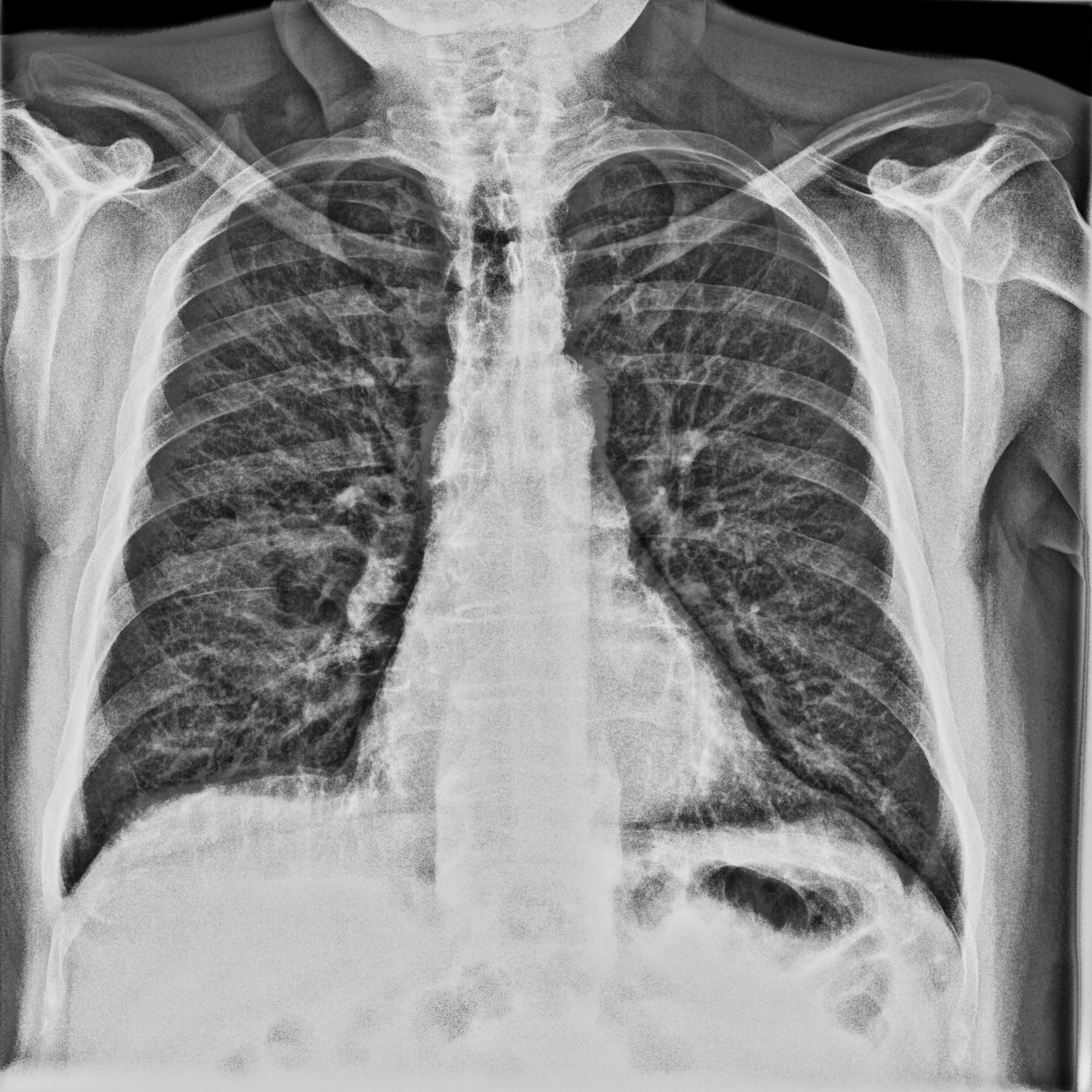} 
 \includegraphics[width=0.24\textwidth]{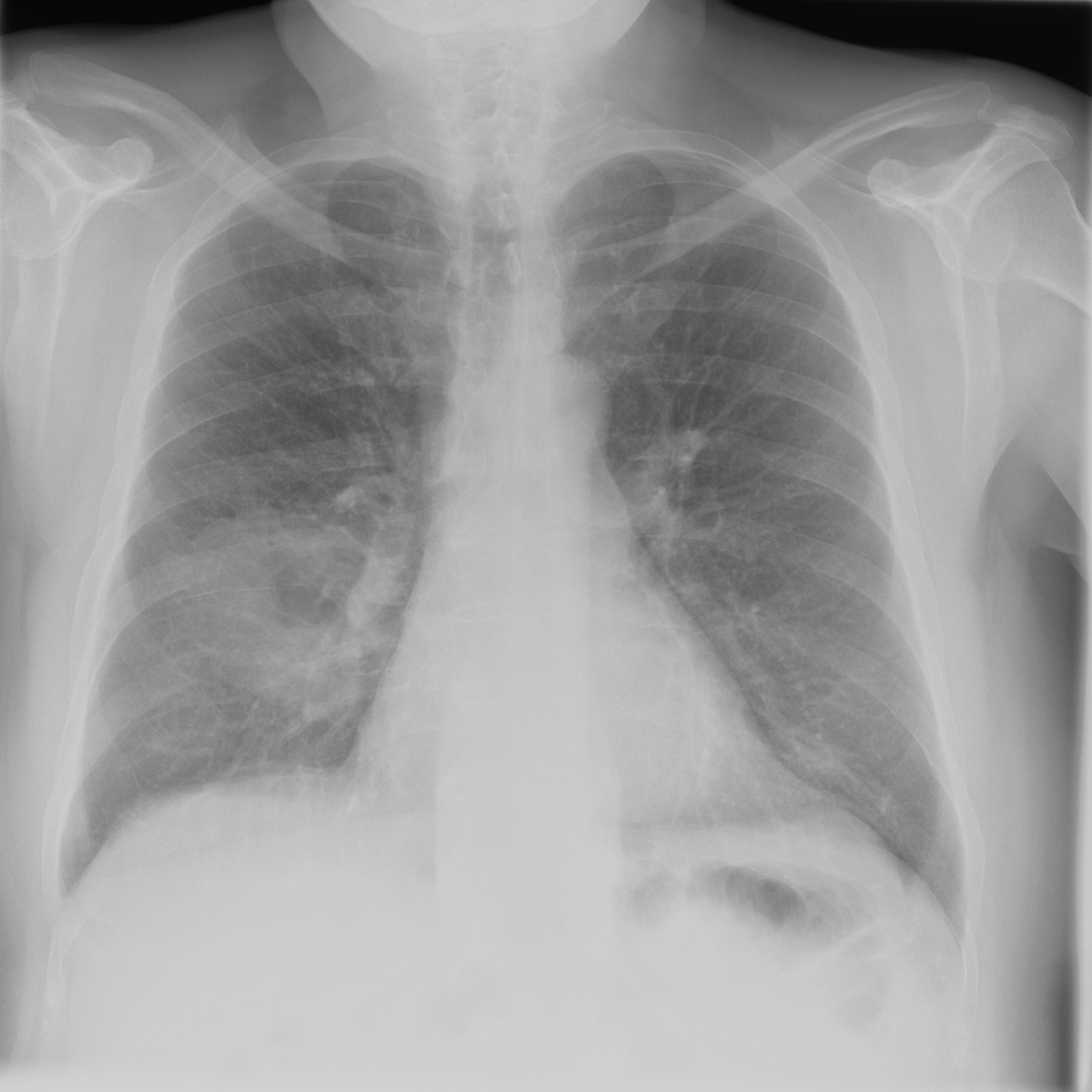}
\includegraphics[width=0.24\textwidth]{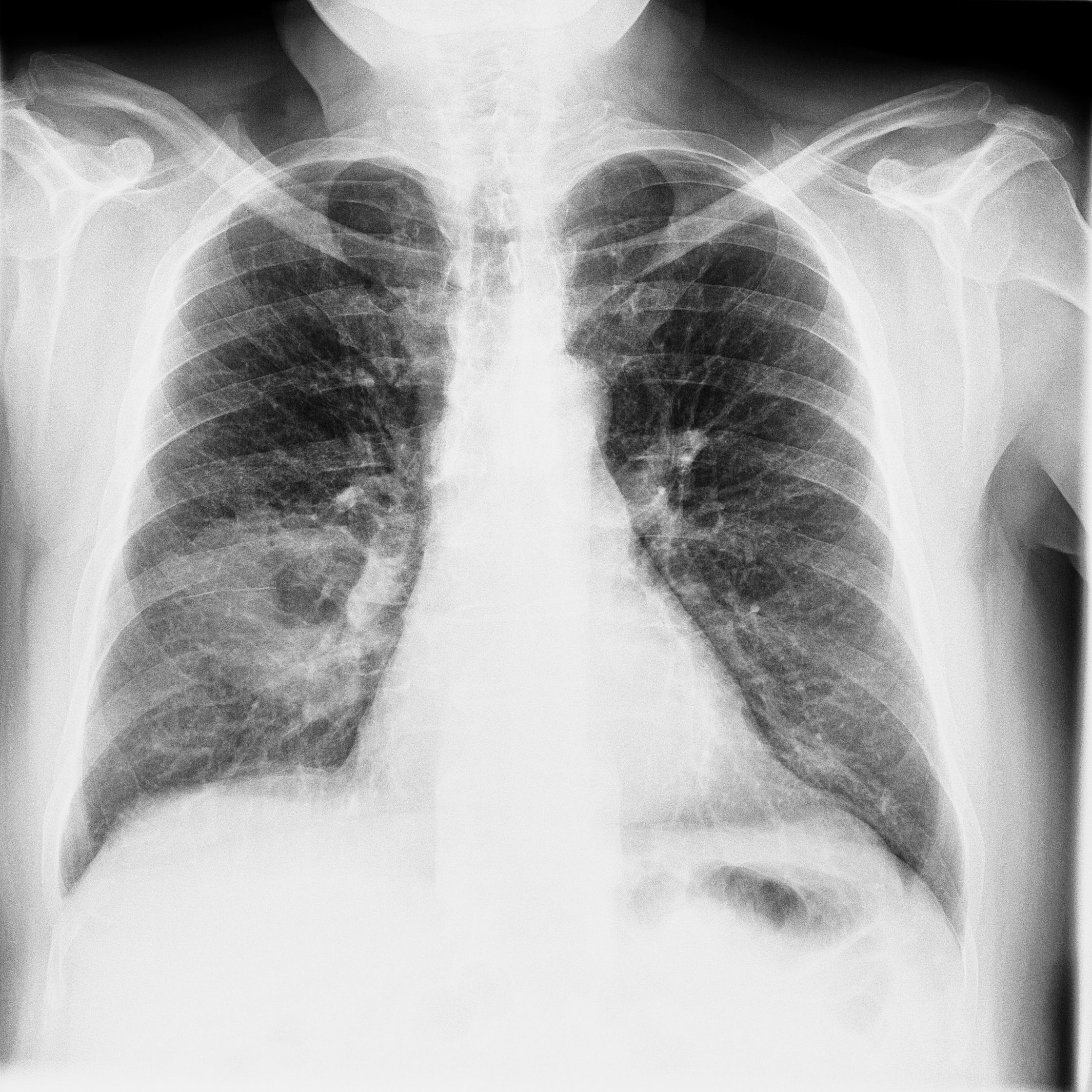}
  \caption{\small Chest X-ray scans with different kinds of post-processing. The image on the left serves as the reference, the other images show lower visual quality.}
  \label{xray1}
\end{figure}

\vspace{-0.15cm}
For the last experiment, we use posteroanterior chest radiographs that were acquired on two imaging systems (Discovery XR656 HD models, GE Healthcare, USA) at Cambridge University Hospitals NHS Trust. Each scanner had previously been set up with different default post-processing parameters (chosen by local radiologists following a subjective assessment), yielding the reference images. Additional images, serving as real-life examples of lower quality, were produced for each radiographic exposure using multiple different post-processing settings, see examples in Figure \ref{xray1}. The post-processing was applied in the hospital directly on the scanner itself by adjusting parameters in the framework provided, including brightness, overall and tissue contrast, edge enhancement, noise reduction, and local contrast enhancement. In total, the data set contains $444$ reference images and $2018$ post-processed images that were rated by $3$ consultant radiologists, $1$ trainee radiologist, and $1$ senior reporting radiographer. Each expert was asked to rate all post-processed images from $1$ (very poor), $2$ (poor), $3$ (good) to $4$ (very good) for general diagnostic purposes in comparison to the reference without changing the contrast or luminance of displayed images. 

\vspace{-1cm}

\section{Results and Discussion}
\vspace{-0.2cm}
\begin{table*}[h!]
\centering
%\begin{tabular}{lrrrr}
\resizebox{0.9\columnwidth}{!}{%
\begin{tabular}{p{0.18\linewidth}p{0.2\linewidth}p{0.2\linewidth}p{0.02\linewidth}p{0.2\linewidth}p{0.2\linewidth}}
\toprule
  & \multicolumn{2}{c}{Grayscale Natural Images} & & \multicolumn{2}{c}{Medical Images} \\
\midrule
 \textit{Full-Reference} & LIVE & LIVE$_{Multi}$ &  & Photoacoustic & Chest X-ray \\
\midrule
PSNR & 0.87 / 0.71 & 0.74 / 0.56 & & 0.71 / 0.54 & 0.66 / 0.48 \\
SSIM & 0.88 / 0.72 & 0.67 / 0.49 & & 0.62 / 0.49 & 0.70 / 0.50 \\ 
SSIM$^*$ & 0.88 / 0.71 & 0.67 / 0.49 & & 0.62 / 0.49 & 0.70 / 0.50 \\
%PSNR$^*$ & 0.87 / 0.71 & 0.74 / 0.56 & & & 0.51 / 0.39 & 0.66 / 0.48 \\
MS-SSIM & 0.91 / 0.77 & 0.88 / 0.70 & & \textbf{0.81/0.65}& 0.80 / 0.58 \\
MS-SSIM$^*$ & 0.91 / 0.76 & 0.88 / 0.71 & & \textbf{0.84/0.68} & 0.79 / 0.57 \\
IW-SSIM & 0.92 / 0.79 & \textbf{0.93/0.77 }& &  0.79 / 0.63 & 0.72 / 0.52 \\
DISTS & 0.91 / 0.76 & 0.75 / 0.56 & & 0.70 / 0.56 & 0.77 / 0.54 \\
DISTS$^*$ & 0.91 / 0.76 & 0.74 / 0.56 & & 0.69 / 0.55 & 0.77 / 0.55 \\
DSS & 0.92 / 0.78 & 0.91 / 0.74& & 0.74 / 0.58 & 0.68 / 0.50 \\
FSIM & \textbf{0.93/0.80 }& 0.92 / 0.75 &  & 0.78 / 0.62 & 0.79 / 0.56 \\
GMSD & \textbf{0.92/0.79 }& 0.91 / 0.74 & & 0.77 / 0.60 & 0.82 / 0.61 \\
HaarPSI & \textbf{0.93/0.79 }& \textbf{0.92/0.76} & & \textbf{0.81/0.65} & \textbf{0.83/0.61}\\
LPIPS$_{Alex}^*$ & 0.90 / 0.75 & 0.77 / 0.59  & & 0.76 / 0.59 & \textbf{0.82/0.62}\\
MDSI & 0.92 / 0.78 & \textbf{0.92/0.76} & & 0.68 / 0.51 & 0.76 / 0.53 \\
VIF$^*$ & 0.85 / 0.68 & 0.90 / 0.72  & & 0.53 / 0.40 & 0.63 / 0.43 \\
VSI & 0.91 / 0.77 & 0.89 / 0.71  & &  0.03 / 0.01 & \textbf{0.83/0.62} \\
\midrule 
\textit{No-Reference}\\
BRISQUE & 0.92 / 0.78 & 0.46 / 0.33 & & 0.59 / 0.44 & 0.05 / 0.03 \\
NIQE & 0.88 / 0.71 & 0.75 / 0.57  & &  0.53 / 0.39 & 0.44 / 0.30 \\
PAQ-2-PIQ$^*$  & 0.76 / 0.57 & 0.86 / 0.68  & & 0.20 / 0.14 & 0.59 / 0.41 \\
\bottomrule
\end{tabular} 
}
\caption{\small SRCC/KRCC of all tested IQA measures and the mean of the rated images' z-scores, described in Section \ref{methods} in Experiment $1, 3, 4$. The top $3$ performers have been printed in bold for each data set. Measures marked with $^*$ have been computed with implementations provided by the authors in Python, for all other measures the provided MATLAB implementations were used. PSNR and HaarPSI provided in both implementations identical results. \vspace{-0.5cm}}
\label{results}
\end{table*}

\begin{figure}
  \centering
\includegraphics[width=0.9\textwidth]{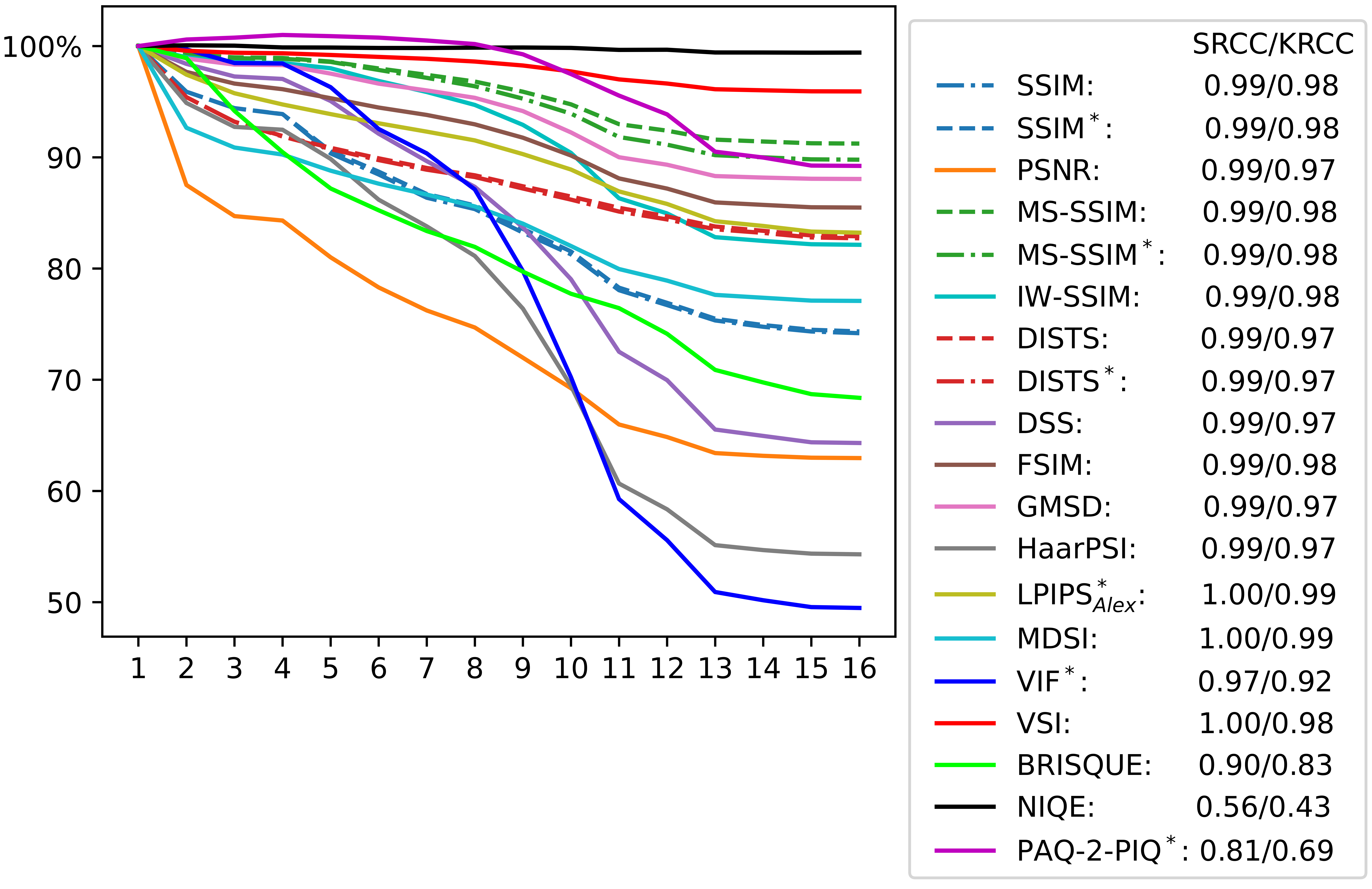} 
\caption{\small IQA comparison of decreasing MRI reconstruction quality through an increase in acceleration factor (1 to 16). All tested FR measures correctly identify a decrease in quality, two of the tested NR measures (NIQE and PAQ-2-PIQ) struggle to identify the quality loss accurately. The SRCC/KRCC values between the measures and the acceleration categories show corresponding behavior. \vspace{-0.5cm}}
  \label{mriplot}
\end{figure}

The question of generalizability across different datasets and target spaces is still an open problem for IQA measures. To gain more insights regarding medical images we conducted $4$ independent experiments, where novel quality ratings were acquired for $4$ data sets and $1$ data set was designed to test the ability to identify obvious quality decrease. For the assessment we chose IQA measures that are commonly used across image modalities and tasks. The results of Experiment $2$ (see Figure \ref{mriplot}) show that all tested FR measures were able to pass the sanity check, i.e.~they were able to identify the distinct quality decrease of the prepared MRI images, confirmed through the descending mean quality value as well as the high SRCC/KRCC values related to the acceleration categories. Two of the tested NR measures, NIQE and PAQ-2-PIQ, struggled to correctly identify the decreasing image quality. \\
In Table \ref{results} we show the SRCC and KRCC values between the tested IQA measures and the $4$ manually rated data sets from Experiments $1, 3$ and $4$. 
The commonly used measures PSNR and SSIM yield relatively low correlation values in all tasks, especially regarding the medical imaging data sets. Outstanding behavior is shown by HaarPSI, which is based on Haar wavelet representations, being among the top 3 performers for all tested data sets. Generally, the correlation values for the natural images are higher than for the medical images, indicating that improvement is still needed in the medical domain and tailored measures should be employed for specified tasks. The reasons for that are manifold. On the one hand, most of the tested measures have been developed and calibrated for natural images, and on the other hand, medical imaging tasks are often very complex or ask for specific quality information. Heavily depending on the task, different image features might be more or less important, such as sharpness, contrast or homogeneity. \\
The tested FR measures yield higher correlation coefficients than the NR measures, which is not surprising as FR assessment uses more information. In the NR setting, recently, measures tailored towards the assessment of specific medical imaging problems have been introduced, see e.g.~\cite{LEI2022102344,CTmultivendor}, and it is advisable to employ such measures in medical imaging tasks in addition to more generalizable measures. Using unsuitable IQA measures might give incorrect conclusions about novel introduced algorithms, not necessarily favoring the most adequate method. Therefore, a change of research culture and extensive research in the area of IQA suitability is needed, including: (1) the development of further task-based measures; (2) the adaption of existing measures (e.g.~through calibration regarding medical images rather than natural images); (3) sharing of medical data with IQA expert ratings (such as the recently introduced CT IQA challenge data set \cite{LEE2025103343}); and (4) more extensive comparison studies. \\
This study explored the correlation between manual expert image ratings and the values provided by common IQA measures in multiple experiments. The results show a trend for the studied medical images and, following previous studies, we conclude that commonly used IQA measures, such as PSNR and SSIM, are not necessarily a suitable fit for medical images and that HaarPSI shows high potential for generalizability also in the medical domain. However, medical imaging problems differ highly and to ensure the suitability of assessment measures for other modalities and tasks, additional analyses have to take place.
\vspace{-0.3cm}
\section{Code and Data Availability}\label{sharing}
\vspace{-0.2cm}
The Python code of the experiments, including the implementations of the Python based IQA measures, is available on GitHub (\url{https://github.com/ideal-iqa/iqa-eval}), as well as the raw csv file containing the $5$ annotations for the grayscale LIVE data of Experiment 1. The corresponding original image data sets are publicly available and may be requested from the LIVE repository (\url{https://live.ece.utexas.edu/research/quality/}). Experiment 2 relies on data from the fastMRI brain dataset, to be requested at \url{https://fastmri.med.nyu.edu/}. The photoacoustic images and annotations in Experiment 3 are available on Zenodo (\url{https://doi.org/10.5281/zenodo.13325197}). The X-Ray image data and annotations used in Experiment 4 are in progress to be made available in a managed way in accordance with the ethical agreements of the acquired clinical data. 
\vspace{-0.3cm}
\section{Conclusion}
\vspace{-0.2cm}
We systematically test the ability of common FR and NR IQA measures to assess the quality of medical images and observe that the measures correlate less with expert assessment of the medical images compared to the grayscale natural images. Even though PSNR and SSIM are commonly used across fields, they yield relatively low results. This confirms previous studies, showing that these two measures are not necessarily a beneficial choice to assess medical imaging tasks, and emphasizing the need to employ task-tailored IQ measures more often. On the other hand, HaarPSI showed exceptional behavior regarding generalizability, suggesting that it may act as a robust measure of quality across domains to complement task-tailored measures. FSIM, GMSD and MS-SSIM also showed relatively robust successful behavior across the data sets; LPIPS and IW-SSIM succeeded in most tasks. \\
In summary, we presented a comprehensive study to assess the adequacy of common IQA measures in medical imaging tasks including rated chest X-ray scans and photoacoustic images, which are made publicly available. We show the dire need for better suited IQA measures and hope to provoke more research in this direction, including extensive suitability studies, adaption of existing measures for medical images as well as further implementation of task-tailored measures. 
%
% ---- Bibliography ----
%
% BibTeX users should specify bibliography style 'splncs04'.
% References will then be sorted and formatted in the correct style.
%

\tiny
\bibliographystyle{splncs04}
\bibliography{bib1}

\end{document}